\documentclass[12pt]{article}
\title{Supersymmetric renormalization prescription  in N = 4 super-Yang--Mills theory}
\usepackage{mathrsfs}
\usepackage{amsmath}

\global\arraycolsep=1pt
\usepackage[colorlinks=true,backref=true,linkcolor=black,anchorcolor=black,citecolor=black,filecolor=black,menucolor=black,pagecolor=black,urlcolor=black]{hyperref}
\usepackage[Symbol]{upgreek}
\usepackage{bbm}
\usepackage{dsfont}
\usepackage{amssymb}
\usepackage{textcomp}
\newcommand{\ba}{/ \hspace{-1.2ex}} 
\newcommand{\baa}{/ \hspace{-1.4ex}}
\newcommand{\baaa}{\, / \hspace{-1.6ex}}

\newcommand{\Scal}[1]{\biggl ({#1} \biggr )}
\newcommand{\scal}[1]{\bigl ({#1} \bigr )}
\def\bea{\begin{eqnarray}}
\def\eea{\end{eqnarray}}
\def\be{\begin{equation}}
\def\ee{\end{equation}}
\newcommand{\CR}{\nonumber \\*}

\newcommand{\petit}{\fontsize{8pt}{7pt}\selectfont}
\newcommand{\norme}{\fontsize{12pt}{14.5pt}\selectfont}

\newcommand{\trace}{\hbox {Tr}~}

\DeclareMathAlphabet{\mathpzc}{OT1}{pzc}{m}{it}
\def\s{\,\mathpzc{s}\,}
\def\a{{\scriptscriptstyle (\mathpzc{s})}}
\def\q{{{\scriptscriptstyle (Q)}}}
\def\qs{{\scriptscriptstyle (Q\mathpzc{s})}}

\def\Q{{\mathcal{S}_{\q}}}
\def\hQ{{\mathcal{S}^{ \scriptscriptstyle \rm ext}_{\q}}}
\newcommand\phis[1]{{\phi^\a_{#1}}}
\newcommand\phiq[1]{{\phi^{\q}_{#1}}}
\newcommand\phiqs[1]{{\phi^{{\qs}}_{#1}}}
\def\blambdas{{\overline{\lambda}^\a}}
\def\blambdaq{{\overline{\lambda}^{\q}}}
\def\blambdaqs{{\overline{\lambda}^{{\qs}}}}
\def\lambdaq{{\lambda^{\q}}}
\def\lambdaqs{{\lambda^{{\qs}}}}
\def\Omegas{{\Omega^\a}}
\def\Omegaq{\Omega^{\q}}
\def\Omegaqs{\Omega^{{\qs}}}
\newcommand\As[1]{{A^\a_{#1}}}
\newcommand\Aq[1]{A^{\q}_{#1}}
\newcommand\Aqs[1]{A^{{\qs}}_{#1}}
\def\cq{c^{{\q}}}
\def\muq{\mu^{{\q}}}

\def\S{{\mathcal{S}_\a}}

\def\F{\mathscr{F}}
\def\G{{\mathcal{G}}}

\def\x{{\mathrm{x}}}

\def\dyx{{\rm \scriptscriptstyle sub}}

\usepackage[colorlinks=true,backref=true,linkcolor=black,anchorcolor=black,citecolor=black,filecolor=black,menucolor=black,pagecolor=black,urlcolor=black]{hyperref}

\usepackage[Symbol]{upgreek}
\usepackage{caption}
\usepackage{bbm}
\usepackage{dsfont}
\usepackage{amssymb}
\usepackage{textcomp}
\def\bea{\begin{eqnarray}}
\def\eea{\end{eqnarray}}
\def\be{\begin{equation}}
\def\ee{\end{equation}}

\newcommand\vo[1]{{\, ^{{ \scriptscriptstyle #1}} \hspace{-0.8mm} v}}
\def\k{{\rm\scriptscriptstyle K}}
\def\bps{{\rm \scriptscriptstyle C}}
\def\uk{{\, ^\k \hspace{-0.7mm} u}}
\def\ub{{\, ^\bps \hspace{-0.7mm} u}}

\def\susy{{\delta^{\mathpzc{Susy}}}}

\def\N{\mathcal{N}}

\begin{document}
\allowdisplaybreaks[1]
\renewcommand{\thefootnote}{\fnsymbol{footnote}}
\def\corr{$\spadesuit $}
\def\trefle{ $\clubsuit$}
\begin{titlepage}
\begin{flushright}
CERN-PH-TH/2006-196
\end{flushright}
\begin{center}
{{\Large \bf
Supersymmetric renormalization prescription  in $\mathcal{N}=4$
super-Yang--Mills theory
}}
\lineskip .75em
\vskip 3em
\normalsize
{\large Laurent Baulieu\footnote{email address: baulieu@lpthe.jussieu.fr} and 
Guillaume Bossard\footnote{email address: bossard@lpthe.jussieu.fr}\\
$^{* }$\it Theoretical Division CERN, { CH-1211 Geneva  23, Switzerland
}
\\
$^{*\dagger}$ {\it LPTHE, CNRS, Universit\'es   Pierre et Marie Curie et Denis~Diderot\\

4 place Jussieu, F-75252 Paris Cedex 05, France}
\\}

\vskip 1 em
\end{center}
\vskip 1 em
\begin{abstract}

\end{abstract}
Using the shadow dependent  decoupled Slavnov--Taylor identities associated to gauge invariance and supersymmetry, 
we discuss the  renormalization of   the  $\N=4$ super-Yang--Mills theory and of  its coupling to gauge-invariant  operators. We specify the method for    the determination of non-supersymmetric counterterms that are needed to maintain supersymmetry.
\end{titlepage}
\renewcommand{\thefootnote}{\arabic{footnote}}
\setcounter{footnote}{0}



\renewcommand{\thefootnote}{\arabic{footnote}}
\setcounter{footnote}{0}



\section{Introduction}
Non-linear aspects of supersymmetry and   the non-existence 
of a
supersymmetry-preserving regulator make the renormalization of
supersymmetric theories a subtle task. Whichever is the choice of
regularization,  we expect  non-supersymmetric counterterms  for 
maintaining supersymmetry at the renormalized level. A very
effective regularization of UV divergences of super-Yang--Mills
theories, called dimensional reduction, was introduced quite early by Siegel \cite{siegel}. Whether this regularization holds true at all orders in perturbation theory was questioned in \cite{consistence}. With suitable improvements, its compatibility with 
the quantum action principle was shown in \cite{reduction}. In fact, this regularization cannot preserve supersymmetry beyond 3-loop
order \cite{brise}, which    implies the introduction of
non-supersymmetric counterterms for  4-loop computations.
 As another complication, the renormalizable Lorentz covariant gauge conditions
(Landau--Feynman-type gauges) break supersymmetry. This breaking of a global
symmetry   is analogous to that of the Lorentz invariance
by axial or Coulomb gauges for the ordinary Yang--Mills theories, but it 
 is more intricate, because supersymmetry is realized non-linearly. This question was addressed   by Dixon \cite{dixon,white}, who completed the ordinary BRST symmetry transformations
for gauge invariance by adding  supersymmetry transformations, whose supersymmetry   parameter is a commuting constant spinor. The ``enlarged BRST symmetry'' determines a  Slavnov--Taylor identity. It was shown,  in a series of papers by St\"{o}ckinger et al., that this process allows    the  determination,  order by order in perturbation theory,  of non-invariant counterterms   \cite{becchi} that restore supersymmetry covariance of Green functions in the $\N=1$ models \cite{stockinger}. The unusual feature that occurs is that Feynman rules depend on the parameter of supersymmetry, but it is 
advocated
 that observables
  do not depend on it. 
This method has
a conceptual backlash. 
To define the ``enlarged BRST symmetry'', Dixon changed the transformation
law of the Faddeev--Popov ghost (to achieve nilpotency of the
``enlarged BRST transformations''). 
But then, the BRST equation of the Faddeev--Popov ghost loses
its geometrical meaning. Moreover, observables are not defined as they should be, from the
cohomology of the BRST differential, since, in this case, they would be
reduced to   supersymmetry scalars.
They must be  introduced as  gauge-invariant  functionals of physical fields, which are  well defined classically,  but are   sources of confusion  at the quantum level, because of  their  possible mixing with  non-gauge-invariant  operators.
 In fact, the previous methods are
sufficient to define certain rules for practical perturbative computations,
 but the way they are obtained lacks  the important
feature of relying on a well-funded algebraic construction. The latter  must be  independent of  the renormalization scheme and clearly separates gauge invariance from supersymmetry.

In recent papers, we indicated the possibility  of disentangling  these two invariances,
 for defining the quantum theory, with 
 independent Slavnov--Taylor identities \cite{shadow}.
We introduced new fields, which we
called shadows, not to confuse them with the usual Faddeev--Popov ghosts. The
advantage of doing so is as follows. The  obtained     pair of differential operators  allows us     to define the two  Slavnov--Taylor operators that characterize the gauge-fixed BRST-invariant supersymmetric quantum field theory, while   the Faddeev--Popov ghost keeps the same  geometrical  interpretation  as in the ordinary Yang--Mills theory. Observables are  defined
by the cohomology of the BRST differential Slavnov--Taylor operator
and their supersymmetry covariance is controlled at the quantum level
by the other Slavnov--Taylor operator for supersymmetry. This will allow for an unambiguous perturbative renormalization of supersymmetric gauge theories.

The shadow
fields are assembled into BRST doublets, and they  do not affect the physical sector. The quantum field theory has an internal 
bigrading, the ordinary ghost number and the new shadow
number. The commuting supersymmetry parameter is understood as an ordinary gauge parameter for the quantum field theory. 
The  prize one has to pay for having shadows is that they generate   a perturbative theory  with  more   Feynman diagrams. If we  consider physical composite operators
that mix through renormalization with BRST-exact operators, we  have  in
principle to consider the whole set of fields in order to compute the
supersymmetry-restoring non-invariant counterterms. For certain  ``simple"  Green
functions, which cannot mix  with  BRST-exact composite
operators, there exist gauges  in which some of the additional fields can be
integrated out, in a way that justifies the work of St\"{o}ckinger et
al. in the $\N=1$ theories. By doing this elimination, we  lose the geometrical
meaning, but we  may gain  in   computational simplicity.\footnote{We do not
exclude the possibility   of  also    reducing  the set of fields in
the general case, including observables that mix with BRST-exact
operators through renormalization, but further investigations are
needed in order to establish this statement. }

In  the conformal phase of the $\N =4$ super-Yang--Mills theory, the observables
are usually defined as   correlation functions of gauge-invariant  operators. 
%
The aim of this paper is to  discuss  their quantum definition and the   methodology that is needed to non-ambiguously  compute non-invariant counterterms and  maintain supersymmetry. In fact,  our results apply to  the renormalization of   all  supersymmetric theories.

\section{Shadow fields and supersymmetry Slavnov--Taylor identities}

\subsection{Action and symmetries}
The physical fields of   the $\N=4$ super-Yang--Mills theory  in $3+1$ dimensions are the gauge field
$A_\mu$ the $SU(4)$-Majorana spinor $\lambda$,  and the six scalar fields
$\phi^i$ in the vector representation of $SO(6) \sim SU(4)$. They are
all in the adjoint representation of a compact gauge group that we
will suppose simple. The classical action is uniquely determined by $Spin(3,1)\times SU(4)$, supersymmetry and gauge invariance. It reads
\be S \equiv \int d^4 x \trace \Bigl(- \frac{1}{4} F_{\mu\nu} F^{\mu\nu} -
\frac{1}{2} D_\mu \phi^i D^\mu \phi_i + \frac{i}{2}
\scal{\overline{\lambda} \baaa D \lambda } - \frac{1}{2} \scal{
\overline{\lambda} [ \phi, \lambda]} - \frac{1}{4}
[\phi^i,\phi^j][\phi_i,\phi_j] \Bigr) \ee
with $\phi \equiv \phi^i \tau_i $ and the supersymmetry transformations $\susy$ \footnote{The parameter $\epsilon$ is a commuting spinor, so that 
 $ \susy^2\approx \delta^{\rm gauge}(\overline{\epsilon} [ \phi - i
\baaa A ] \epsilon) - i (\overline{\epsilon} \gamma^\mu \epsilon)
\partial_\mu$, where $\approx $ stands for the equality modulo equations of motion.}
\begin{gather}
\susy A_\mu = i \scal{\overline{\epsilon} \gamma_\mu \lambda }
\hspace{10mm} \susy \phi^i = - \scal{ \overline{\epsilon} \tau^i
\lambda}
\hspace{10mm} 
\susy \lambda = \scal{ \baa F + i \baaa D \phi + \frac{1}{2} [\phi,
\phi]} \epsilon 
\end{gather}
Out of $\susy$, we  can build an operator $Q$ that also acts on the Faddeev--Popov field 
$\Omega$ and new shadow fields $c,\,\mu$~\cite{shadow}.
 $Q$ acts on all the physical fields as $Q=\susy-\delta^{\rm gauge}(c)$, and we  have
 \begin{gather}
Q c = ( \overline{\epsilon} [ \phi - i
\baaa A ] \epsilon) - c^2 
\hspace{10mm}
Q \Omega = - \mu - [c, \Omega] 
\CR   Q \mu = - [ (
\overline{\epsilon} \phi \epsilon ) , \Omega ] + i (\overline{\epsilon}
\gamma^\mu \epsilon) D_\mu \Omega - [c, \mu]
\end{gather} 
The BRST operator $\s $ is nothing but a gauge transformation
of parameter $\Omega$ on all physical fields, and we have 
\be \s \Omega = - \Omega^2 \hspace{7mm} \s c = \mu \hspace{7mm} \s \mu
= 0 \ee
To define a BRST-exact supersymmetric   gauge-fixing, we introduce the
trivial quartet $\bar \mu ,\, \bar c,\,\bar \Omega,\, b$, with
\begin{gather}\begin{split}
\s \bar \mu &= \bar c \\* Q \bar \mu &= \bar \Omega
\end{split}\hspace{10mm}\begin{split}
\s \bar c &= 0 \\* Q \bar c &= - b
\end{split}\hspace{10mm}\begin{split}
\s \bar \Omega &= b \\ Q \bar \Omega &= - i (\overline{\epsilon}
\gamma^\mu \epsilon) \partial_\mu \bar \mu
\end{split}\hspace{10mm}\begin{split}
\s b &=0 \\* Q b&= i (\overline{\epsilon}
\gamma^\mu \epsilon) \partial_\mu \bar c
\end{split}\end{gather}
On all fields, we  have $
\s^2 = 0,\ Q^2 \approx -i (\overline{\epsilon} \gamma^\mu
\epsilon) \partial_\mu,\ 
\{\s , Q \} = 0
$. We  have the following  renormalizable supersymmetric $\s
 Q$-exact gauge-fixing actions\footnote{Note that power counting forbids a gluino dependence for the argument of the $\s Q$-exact term, and that $Q$ is nilpotent on all the functionals that do not depend on the gluinos.   $\alpha $ is the usual interpolating Landau--Feynman  gauge parameter.}
\be -\s Q \int d^4 x \trace \Bigl( \bar \mu
\partial^\mu A_\mu + \frac{\alpha}{2} \bar \mu b  \Bigr) \ee
By introducing  sources associated to the non-linear  $\s$, $Q$  and $\s Q$ transformations of fields, we  get the following   $\epsilon$-dependent action, which   initiates a BRST-invariant  supersymmetric perturbation theory\footnote{
$M$ is the $32 \times 32$ matrix 
$ M \equiv \frac{1}{2} (\overline{\epsilon} \gamma^\mu \epsilon)
\gamma_\mu + \frac{1}{2} (\overline{\epsilon} \tau_i \epsilon)
\tau^i - \epsilon \overline{\epsilon} $. It occurs 
 because $Q^2$ is a pure derivative only modulo
 equations of motion.
 The 
dimension of $A_\mu,\, \lambda ,\, \phi^i,\, \Omega,\, \bar \Omega ,\,
b,\, \mu,\, \bar \mu,\, c$ and $\bar c$ are respectively $1,\,
\frac{3}{2},\, 1,\, 0,\, 2,\, 2,\, \frac{1}{2},\, \frac{3}{2},\,
\frac{1}{2}$ and $\frac{3}{2}$. Their ghost and shadow numbers are respectively
$(0,0),\,(0,0),\,(0,0),\,
(1,0),\,(-1,0),\,(0,0),\,(1,1),\,(-1,-1),\,(0,1) $ and $(0,-1)$.}
\begin{multline} 
\Sigma \equiv \frac{1}{g^2} S - \int d^4 x \trace \Bigl( b
\partial^\mu A_\mu + \frac{\alpha}{2} b^2 - \bar c
\partial^\mu \scal{D_\mu c + i(\overline{\epsilon} \gamma_\mu \lambda)}
- \frac{i \alpha}{2} (\overline{\epsilon} \gamma^\mu \epsilon) \bar c
\partial_\mu \bar c \\*\hspace{50mm} + \bar \Omega \partial^\mu
D_\mu \Omega - \bar \mu \partial^\mu \scal{D_\mu \mu + [D_\mu \Omega, c] - i
(\overline{\epsilon} \gamma_\mu [\Omega, \lambda])} \Bigr) \\*
+ \int d^4 x \trace \biggl( \As{\mu} D^\mu \Omega + \blambdas [\Omega,
\lambda] - \phis{i} [\Omega , \phi^i] + \Aq{\mu} Q A^\mu - \blambdaq Q \lambda + \phiq{i} Q \phi^i \\* 
+ \Aqs{\mu} \s Q A^\mu - \blambdaqs \s Q \lambda + \phiqs{i} \s Q
\phi^i + \Omegas \Omega^2 -
\Omegaq Q \Omega - \Omegaqs \s Q \Omega \\* - \cq Q c + \muq Q \mu
+ \frac{\, g^2}{2} ( \blambdaq - [\blambdaqs , \Omega ]) M ( \lambdaq -
[\lambdaqs , \Omega ]) \biggr)
\end{multline}

Because of the $\s$ and $Q$ invariances, the action is invariant under both 
Slavnov--Taylor identities defined in \cite{shadow}, which are  associated
respectively to gauge and supersymmetry invariance,
$ \S(\Sigma) = \Q(\Sigma) = 0$. The supersymmetry Slavnov--Taylor operator is\footnote{ The linearized Slavnov--Taylor operator  ${\Q _{|\Sigma}}$   \cite{shadow} verifies 
  ${\Q _{|\Sigma}} ^2 =-i (\overline{\epsilon} \gamma^\mu
\epsilon) \partial_\mu$, which solves in practice the fact that $Q^2$
is a pure derivative only modulo equations of motion.}
\begin{multline}
\Q(\F) \equiv \int d^4 x \trace \biggl( \frac{\delta^R \F}{\delta A^\mu}
\frac{\delta^L \F}{\delta \Aq{\mu}} + \frac{\delta^R \F}{\delta \lambda}
\frac{\delta^L \F}{\delta \blambdaq}+ \frac{\delta^R \F}{\delta \phi^i}
\frac{\delta^L \F}{\delta \phiq{i}} + \frac{\delta^R \F}{\delta c}
\frac{\delta^L \F}{\delta \cq} +\frac{\delta^R \F}{\delta
\mu}\frac{\delta^L \F}{\delta \muq} \biggr .\\* + \frac{\delta^R \F}{\delta \Omega}
\frac{\delta^L \F}{\delta \Omegaq} - \As{\mu}
\frac{\delta^L \F}{\delta \Aqs{\mu}} + \blambdas \frac{\delta^L \F}{\delta
\blambdaqs} - \phis{i} \frac{\delta^L \F}{\delta
\phiqs{i}} + \Omegas \frac{\delta^L \F}{\delta \Omegaqs} - b
\frac{\delta^L \F}{\delta \bar c} + \bar
\Omega \frac{\delta^L \F}{\delta \bar \mu} 
\\*-i (\overline{\epsilon} \gamma^\mu \epsilon)\Bigl(- \partial_\mu
\Aqs{\nu} \frac{\delta^L \F}{\delta \As{\nu}} + \partial_\mu
\blambdaqs \frac{\delta^L \F}{\delta \blambdas}- \partial_\mu
\phiqs{i} \frac{\delta^L \F}{\delta \phis{i}} + \partial_\mu
\Omegaqs \frac{\delta^L \F}{\delta \Omegas} - \partial_\mu \bar c \frac{\delta^L \F}{\delta b} + \partial_\mu \bar \mu \frac{\delta^L \F}{\delta \bar \Omega} \Bigr . \\* + \Aq{\nu} \partial_\mu 
A^\nu + \blambdaq \partial_\mu \lambda + \phiq{i} \partial_\mu
\phi^i + \Omegaq
\partial_\mu \Omega+ \cq \partial_\mu c + \muq \partial_\mu \mu \Bigr) \biggr)
\label{slavnovQ}
\end{multline}

\subsection{ Observables}
The   observables of  the $\N = 4$ super-Yang--Mills theory  in the conformal phase are Green functions of
local operators in the cohomology of the BRST linearized Slavnov--Taylor operator $\S_{|\Sigma}$.  From this definition,  these      Green functions   are independent of the gauge parameters of the action, including $\epsilon$.  Classically, they are represented by  gauge-invariant  polynomials of the physical fields \cite{shadow,henneaux}. We  introduce classical sources $u$ for all these operators.  We must generalize the supersymmetry Slavnov--Taylor identity for the extended local action that   depends on  these sources. Since the supersymmetry algebra does not close off-shell,   other sources $v$, coupled to unphysical $\S_{|\Sigma}$-exact operators, must also be introduced. We define the following field and source combinations $\varphi^*$
\be\begin{split}
A^*_\mu &\equiv \Aq{\mu} - \partial_\mu \bar c - [\Aqs{\mu} - \partial_\mu \bar \mu , \Omega] \\*
\phi^*_i &\equiv \phiq{i} - [\phiqs{i} , \Omega]
\end{split}\hspace{10mm}\begin{split}
c^* &\equiv \cq - [\muq,\Omega] \\*
\lambda^* &\equiv \lambdaq - [\lambdaqs , \Omega] 
\end{split}\ee
They verify $\S_{|\Sigma} \varphi^* = - [\Omega, \varphi^*]$. The  collection  of local operators coupled to the $v$'s  is made of all possible gauge-invariant (i.e. $\S_{|\Sigma}$-invariant) 
 polynomials in the physical fields and   the $\varphi^*$'s. These operators have ghost number zero, and their shadow number is negative, in contrast with the physical gauge-invariant  operators, which have shadow number zero.

The relevant action is thus $\Sigma[u,v] \equiv \Sigma + \Upsilon[u,v]$, with 
\begin{multline}
\Upsilon[u,v] \equiv \int d^4 x \biggl( u_{ij} \frac{1}{2}
\trace \phi^i \phi^j + u^\alpha_i \trace \phi^i \lambda_\alpha +
u_{ijk} \frac{1}{3}\trace \phi^i \phi^j \phi^k \\*+ \uk_{ij}^\mu \trace \scal{ i \phi^{[i}
D_\mu \phi^{j]} + \frac{1}{8} \overline{\lambda} \gamma_\mu \tau^{ij}
\lambda} + \uk^{\mu\nu}_i \trace \scal{ F_{\mu\nu} \phi^i
-\frac{1}{2} \overline{\lambda} \gamma_{\mu\nu} \tau^i \lambda} + \uk^5_\mu \frac{1}{2} \trace \overline{\lambda}
\gamma_5 \gamma^\mu \lambda \\*+ \ub_{ijk} \trace\scal{\frac{1}{3}
 \phi^{[i} \phi^j \phi^{k]} + \frac{1}{8} \overline{\lambda} \tau^{ijk}
 \lambda } + \ub_{ij}^\mu \trace \scal{ i \phi^{[i}
D_\mu \phi^{j]} - \frac{1}{4} \overline{\lambda} \gamma_\mu \tau^{ij}
\lambda} \\* + \ub^{\mu\nu}_i \trace \scal{ F_{\mu\nu} \phi^i +\frac{1}{4}
\overline{\lambda} \gamma_{\mu\nu} \tau^i \lambda} + u_{ij}^\alpha
\trace \phi^i \phi^j \lambda_\alpha + i u_i^{\mu\, \alpha} \trace D_\mu
\phi^i \lambda_\alpha + u^{\mu\nu\, \alpha} \trace F_{\mu\nu}
\lambda_\alpha + \cdots \\*
+ v_i^\alpha \trace
\phi^i \lambda^*_\alpha + v^{\alpha\beta} \trace \lambda_\alpha
\lambda^*_\beta + v_i^\mu \trace \phi^i A^*_\mu + v_{ij} \trace \phi^i
\phi^{*\, j} + i v^{\mu\, \alpha}_i \trace D_\mu \phi^i \lambda^*_\alpha \hspace{9mm}\\*+
\vo{0}^\alpha_i \trace \lambda_\alpha \phi^{*\, i} + i v^{\mu
 \alpha\beta} \trace D_\mu \lambda_\alpha \lambda^*_\beta + i 
v^\mu_{ij} \trace D_\mu \phi^i \phi^{*\, j} + i \vo{-1}_i^{\mu\alpha} \trace D_\mu
\lambda_\alpha \phi^{*\, i} + \cdots \biggr)
\end{multline} 
Here, the  $\cdots$ stand for all   other analogous  operators.

The Slavnov--Taylor operator $\Q$ can be generalized into a new one, $\hQ$, by addition of terms that are linear in the functional derivatives with respect to the sources $u$ and $v$, in such a way that  
\be\label{ext}
\hQ(\Sigma[u,v])  =\Q(\Sigma) + \hQ_{|\Sigma} \Upsilon + \int d^4
x \trace \biggl( \frac{\delta^R \Upsilon}{\delta A^\mu}
\frac{\delta^L \Upsilon}{\delta A^*_\mu} + \frac{\delta^R
\Upsilon}{\delta \lambda} \frac{\delta^L \Upsilon}{\delta
\overline{\lambda}^*}+ \frac{\delta^R \Upsilon}{\delta \phi^i}
\frac{\delta^L \Upsilon}{\delta \phi^*_i} \biggr) =0
\ee
Indeed,   if we were to compute $\Q(\Sigma[u,v])$ without taking
into account the transformations of the sources $u$ and $v$, the breaking of the Slavnov--Taylor identity
would be a local functional linear in the set of gauge-invariant  local
polynomials in the physical fields, $A^*_\mu$, $c^*$,
$\phi^*_i$ and $\lambda^*$.

Eq.~(\ref{ext}) defines the 
transformations  $ \hQ_{|\Sigma}$   of   the   sources $u$ and $v$. 
Simplest examples for the transformation laws of   the $u$'s are for instance 
\bea
\hQ_{|\Sigma} u_{ij} &=& - i[\gamma^\mu \tau_{\{i} \epsilon]_\alpha \partial_\mu
u^\alpha_{j\}} + \partial_\mu \partial^\mu v_{\{ij\}} + 2 u_{\{i|k}
{v_{j\}}}^k + 2
u^\alpha_{\{i} v_{j\}\alpha} - i \partial_\mu ( u_{\{i|k} {v^\mu _{j\}}}^k
+ u^\alpha_{\{i} v^\mu_{j\}\alpha}) \CR
\hQ_{|\Sigma} u_i^\alpha &=& [\overline{\epsilon} \tau^j ]^\alpha \scal{ u_{ij} -
 i \partial_\mu ( \uk^\mu_{ij} + \ub^\mu_{ij} )} - 2i [\overline{\epsilon} \gamma_\mu
]^\alpha \partial_\nu ( \uk_i^{\mu\nu} + \ub_i^{\mu\nu} )+ i {[\gamma^\mu]_\beta}^\alpha
\partial_\mu v^\beta_i \CR
& & \hspace{10mm} - u_{ij} \vo{0}^{j\alpha} - u^\alpha_j {v_i}^j
+ u^\beta_i {v^\alpha}_\beta + u^{\alpha\beta} v_{i\beta} +
i \partial_\mu ( u_{ij} \vo{-1}^{j\mu\alpha} - u^\beta_i v_\beta^{\mu \alpha} )
\eea
These transformations are quite complicated in their most general expression. However, for
many practical computations of     non-supersymmetric 
 local counterterms
, we can consider them at $v=0$. We  define $ Q u \equiv \scal{\hQ_{|\Sigma} u}_{|v=0}$. By using $\susy \Upsilon[u] + \Upsilon[ Q u ] = 0$ we can in fact conveniently compute $Qu$. Notice that $Q$ is not nilpotent on the sources, but we have
the result that $\Upsilon[Q^2 u]$ is a linear functional of the equation of
motion of the fermion $\lambda$.

In \cite{shadow,beta}, we   showed the absence of anomaly and the
stability of the $\N=4$ action $\Sigma$ under renormalization. Thus,
the complete theory involving shadows and ghosts can be renormalized,
in any given regularization scheme, so that supersymmetry and gauge
invariance are preserved at any given finite order. It is a
straightforward and precisely defined process to compute all
observables, provided that a complete set of sources has been
introduced. This lengthy process cannot be avoided because  there  exists no 
regulator that preserves both gauge invariance and
supersymmetry. We must keep in mind that   renormalization generally
mixes physical observables with BRST-exact operators, and a careful
analysis must be done \cite{zuber}.

\section{Enforcement of supersymmetry}

We now turn to the problem of determining non-invariant counterterms, which 
  are necessary to ensure supersymmetry at the quantum level. We
will use the notation of \cite{becchi} for the 1PI correlation
functions, an example of which is
\be \bigl< A^a_\mu (p) \lambda^b_\alpha (k) \phi^{{{\scriptscriptstyle (Q)}}\,
c}_i \bigl> \equiv 
\int d^4 x d^4 y \ e^{i (p x + k y)} \,
\frac{\delta^L\ \delta^L\ \delta^L\ \Gamma}{ \delta
\phi^{{{\scriptscriptstyle (Q)}}\, c\, i}(0) \delta \overline{ \lambda}^{b\, \alpha}(y) \delta A^{a\, \mu}(x)} \label{1PI}\ee
All fields and sources are set equal to zero after the differentiation. Latin letters $a,\, b,\,$... label the index of the gauge Lie algebra. 
In this section we  focus on the   case of  
observables that  do not mix with non-gauge-invariant 
operators.   The following subsection explains how   
computations are simplified in this case.

\subsection{Loop cancellations}

 We  can first  eliminate by Gaussian integration the Faddeev--Popov
ghosts $\Omega,\, \bar\Omega$ against the shadows $\bar\mu$, $\mu$ for   computing   some
observables,  in our  class of linear gauges.

The   antighost Ward identities of  \cite{shadow} 
determine the following dependence of the 1PI generating functional $\Gamma$ in
the fields $\bar \mu,\, \bar \Omega,\, \bar c$ and $b$
\begin{multline} \Gamma[\cdots,    \bar \mu, \bar c, \bar \Omega, b ,
  \As{\mu} , \Aq{\mu} , \Aqs{\mu} ] = \Gamma [\cdots,
  0,0,0,0,\As{\mu}- \partial_\mu \bar \Omega,\Aq{\mu} + \partial_\mu
  \bar c , \Aqs{\mu}  + \partial_\mu \bar \mu]\\ - \int d^4 x \trace
  \scal{ b \partial^\mu A_\mu + \frac{\alpha}{2} b^2 - \frac{i \alpha
    }{2} (\overline{\epsilon} \gamma^\mu \epsilon) \bar c\partial_\mu
    \bar c} \label{antighost}\end{multline}
where the $\cdots$ stand  for  the dependence on all other fields and
sources. Consider  the  generating functional of 1PI  correlation functions   of the subset of fields $\varphi_{\dyx}$ made of the physical
fields, the shadow $c$, $\bar c$ and the sources associated to the $Q$
variations of these fields. The pair of $Q$-doublets
$\Omega,\, \mu$ and $ \bar\mu,\,\bar\Omega$   only appear in the
Feynman diagrams through   propagators and   interactions defined
by the following part of the action
\be \int d^4 x \trace \scal{ \partial^\mu \bar \Omega D_\mu \Omega -
  \partial^\mu \bar \mu D_\mu \mu} \ee
Feynman rules show that   the  fields $\Omega,\, \bar\Omega,\, \mu$ and $\bar \mu$   exactly 
   compensate in closed loops of opposite contributions at least at the regularized level.
   The following Ward identities  imply that this property is maintained after renormalization
\begin{gather}
\bigl< \bar\mu^a (p) \mu^b \bigr> + \bigl< \bar \Omega^a (p) \Omega^c
\bigr> \bigl< \Omega^{\scriptscriptstyle (Q)}_c (p) \mu^b \bigr> = 0 \CR
\bigl< \bar\mu^a (p) A^b_\mu (k)  \mu^c \bigr> + \bigl< \bar \Omega^a
(p) A^b_\mu (k) \Omega^d
\bigr> \bigl< \Omega^{\scriptscriptstyle (Q)}_d (p+k) \mu^c \bigr> = 0
\end{gather}
$\Omegaq$ is the source of the   operator $\mu + [\Omega,c]$ and 
 the term linear in $\mu $ of the
insertion of $[\Omega,c]$ in $\Gamma$ must be zero.  It follows  
that $\bigl< \Omega^{\scriptscriptstyle (Q)}_c (p) \mu^b \bigr>=\delta^b_c$, at any given finite  order of
perturbation theory. The only superficially divergent 1PI Green functions depending 
on  $ \Omega,\, \bar \Omega,\, \mu$ and $\bar \mu$
that must  be considered are $\bigl< \bar\mu^a (p) \mu^b \bigr> = -
\bigl< \bar \Omega^a (p) \Omega^b \bigr>$ and $ \bigl< \bar\mu^a (p)
A^b_\mu (k)  \mu^c \bigr> = - \bigl< \bar \Omega^a (p) A^b_\mu (k)
\Omega^c \bigr>$. We can thus integrate out these fields in all
correlation functions     of the fields $\varphi_{\dyx}$.

After this elimination, 
 the supersymmetry Slavnov--Taylor identity is sufficient   to constrain the   1PI Green functions  of the fields  $\varphi_{\dyx}$  to the same values as   they would have in the complete procedure without the ab-initio elimination of  $\Omega,\, \bar\Omega,\, \mu$ and $\bar \mu$.  
In fact,   the most  general local
functional, which satisfies the supersymmetry
Slavnov--Taylor identity and  only  depends on the fields $\varphi_{\dyx}$,    is the restriction of the most  general local functional that satisfies all the
Ward identities of the theory when  all the other fields  and sources are set equal  
to zero.  This justifies the heuristic  argument  that the  Ward identity of   supersymmetry implies that of    gauge invariance because supersymmetry   transformations close modulo gauge transformations. 
Thus, to compute correlation functions of fields   $\varphi_{\dyx}$, we can use the simplified action
\begin{multline} \label{effective}
\Sigma^{\dyx} \equiv \frac{1}{g^2} S + Q \int d^4 x \trace \Scal{ \bar c
\partial^\mu A_\mu + \frac{\alpha}{2} \bar c b }  \\*
+ \int d^4 x \trace \biggl( A^{{\scriptscriptstyle (Q)}}_\mu   Q A^\mu - \overline{\lambda}^{{\scriptscriptstyle (Q)}} Q \lambda + \phi^{{\scriptscriptstyle (Q)}}_i Q \phi^i -c^{{\scriptscriptstyle (Q)}} Q c  
+ \frac{\, g^2}{2} \overline{\lambda}^{{\scriptscriptstyle (Q)}} M \lambda^{{\scriptscriptstyle (Q)}}  \biggr)
\end{multline}
The ambiguities of the quantum theory are fixed  by the antighost Ward identities for  
$\bar c$ and $b$, and the  following simplified supersymmetry Slavnov--Taylor identity
\begin{multline}
\Q (\F) \equiv \int d^4 x \trace \biggl( \frac{\delta^R \F}{\delta A^\mu}
\frac{\delta^L \F}{\delta A^{{\scriptscriptstyle (Q)}}_\mu} + \frac{\delta^R \F}{\delta \lambda}
\frac{\delta^L \F}{\delta \overline{\lambda}^{{\scriptscriptstyle (Q)}}}+ \frac{\delta^R
\F}{\delta \phi^i} \frac{\delta^L \F}{\delta \phi^{{\scriptscriptstyle (Q)}}_i} +
\frac{\delta^R \F}{\delta c}\frac{\delta^L \F}{\delta c^{{\scriptscriptstyle (Q)}}} \\*
-i (\overline{\epsilon} \gamma^\mu \epsilon) \Bigl( A^{{\scriptscriptstyle (Q)}}_\nu  \partial_\mu
A^\nu + \overline{\lambda}^{{\scriptscriptstyle (Q)}} \partial_\mu \lambda + \phi^{{\scriptscriptstyle (Q)}}_i
\partial_\mu \phi^i +c^{{\scriptscriptstyle (Q)}} \partial_\mu c  \Bigr)
- b \frac{\delta^L \F}{\delta \bar c} + i(\overline{\epsilon} \gamma^\mu
\epsilon) \partial_\mu \bar c \frac{\delta^L \F}{\delta b} \biggr)
\label{slavnov*}
\end{multline}
This  identity  is analogous to that     in
\cite{dixon,white, stockinger}. However,    we now understand that $c$ is not the
Faddeev--Popov ghost, and  that observables must be defined in the enlarged theory.

This simplified   process with less fields  can be  applied  also  for     computing
  1PI correlation functions with    insertions of certain  physical
composite operators (we call them ``simple" operators), as long as these operators  do not mix through
renormalization with BRST-exact operators (which would  imply   computing insertions 
of operators depending on   other fields than the $\varphi_{\dyx}$). At the tree level,  these ``simple"
operators are  all the  gauge-invariant  polynomials in the physical fields
that are in representations of $Spin(3,1) \times SU(4)$ in which there
  exist no BRST-exact operators of the same canonical dimensions
that depend on the antighost $\bar \mu,\, \bar \Omega$ and $\bar c$
only through their derivatives.  Examples of ``simple" operators  are
the local  operators  of canonical dimension $[\mathcal{O}] < 4$  and
the BPS primary operators.

\subsection{Renormalization of the action}


We assume that the  ``restricted"    theory has been renormalized at a given order of perturbation theory, say $n$, by using the best available regularization, namely dimensional reduction, and renormalization conditions such that the supersymmetry Slavnov--Taylor identity and the so-called antighost Ward identities are satisfied.
Within this scheme, finite
gauge-invariant, but not supersymmetric, counterterms must occur after a certain 
  order of perturbation theory.
At a given
order $n$,  the action is thus of the
following form
\begin{multline}\label{baction}
\Sigma^\flat = \frac{1}{g^2} \int d^4 x \trace
\biggl( - \frac{1}{4} F^\flat_{\mu\nu} F^{\flat\, \mu\nu} -
D^\flat_\mu \phi^{\flat\, i} D^{\flat\, \mu} \phi^\flat_i +
\frac{i}{2} \scal{\overline{\lambda}^\flat \baaa D^\flat \lambda^\flat } \\*-
\frac{g_1}{2}\scal{\overline{\lambda}^\flat [ \phi^\flat,
\lambda^\flat]} - \frac{g_2}{4} [\phi^{\flat\, i}, \phi^{\flat\, j}][
\phi^\flat_i, \phi^\flat_j] + h_1 \phi^{\flat\,\{i}
\phi^{\flat\, j\}} \phi^\flat_{\{i} \phi^\flat_{j\}} + h_2
\phi^{\flat\, i} \phi^\flat_i \phi^{\flat\, j} \phi^\flat_j \biggr) \\*+
\int d^4 x \Scal{ g_3 \trace \phi^{\flat\,\{i}
\phi^{\flat\, j\}} \trace \phi^\flat_{\{i} \phi^\flat_{j\}} + h_3
\trace \phi^{\flat\, i} \phi^\flat_i \trace \phi^{\flat\, j}
\phi^\flat_j }\\
+ \int d^4 x \trace \Bigl( - b^\flat \partial^\mu A^\flat_\mu -
\frac{\alpha^\flat}{2} {b^\flat}^2 + \bar c^\flat \partial^\mu \scal{
 D^\flat_\mu c^\flat + i {\rm y}_1 (\overline{\epsilon} \gamma_\mu
 \lambda^\flat)} + {\rm y}_2 \frac{i \alpha^\flat}{2}
(\overline{\epsilon} \gamma^\mu \epsilon) \bar c^\flat \partial_\mu
\bar c^\flat \Bigr) \\
+ \int d^4x \trace \biggl( A^{{{\scriptscriptstyle (Q)}}\flat}_\mu \scal{ i \x_1
(\overline{\epsilon} \gamma^\mu \lambda^\flat) + D^{\flat\, \mu}
c^\flat} - \phi^{{{\scriptscriptstyle (Q)}}\, \flat}_i \scal{ \x_2
(\overline{\epsilon}\tau^i\lambda^\flat ) + [c^\flat, \phi^{\flat\,
i}]} \\* + \overline{\lambda}^{{{\scriptscriptstyle (Q)}}\, \flat} \scal{ - \bigl[ \x_3
\baa F^\flat + i \x_4 \baaa D^\flat \phi^\flat + \frac{\x_5}{2}
[\phi^\flat, \phi^\flat] + h_4 \phi^{\flat\, i}\phi^\flat_i
\bigr]\epsilon + [ c^\flat , \lambda^\flat]} \\*
+ c^{{{\scriptscriptstyle (Q)}}\, \flat} \scal{ - \x_6 (\overline{\epsilon} \phi^\flat
\epsilon) + i \x_7 (\overline{\epsilon} \baaa A^\flat \epsilon) +
{c^\flat}^2 } + \frac{\, g^2}{2} \,\overline{\lambda}^{{{\scriptscriptstyle (Q)}}\, \flat} N
\lambda^{{{\scriptscriptstyle (Q)}}\, \flat} \biggr) 
\end{multline}
The conformal 
property of $\N=4$ implies that the coupling constant is not
renormalized. In this expression, the   index $\flat$   on top of 
a  field $\varphi$ indicates its multiplicative renormalization
by an infinite factor $\sqrt{ Z_ \varphi} $, which  is a Taylor series of
order $n$ in the coupling constant $g^2 $. The
sources $\varphi ^{\scriptscriptstyle (Q)}$ are renormalized by the
inverse factor $1/ \sqrt{ Z_ \varphi }$ as a result of  the  BRST
Slavnov--Taylor identities. The parameters $g_I$,
$h_I$, $\x_I$, ${\rm y}_I$ and the $32\times 32$ symmetric matrix $N$
(quadratic in $\epsilon$)\footnote{$N$ can be parametrized by five
 parameters as follows \vskip -8mm
\be N \equiv {\rm a}_1 (\overline{\epsilon} \gamma^\mu \epsilon) \gamma_\mu +
{\rm a}_2 (\overline{\epsilon} \tau^i \epsilon) \tau_i + 6 {\rm a}_3
(\overline{\epsilon} \gamma^{\mu\nu} \tau^{ijk} \epsilon)
\gamma_{\mu\nu} \tau_{ijk} + 2 {\rm a}_4 (\overline{\epsilon} \gamma_5
\gamma^\mu \tau^{ij} \epsilon) \gamma_5 \gamma_\mu \tau_{ij} +
{\rm a}_5 (\overline{\epsilon} \gamma_5 \tau^i \epsilon) \gamma_5
\tau_i \nonumber \ee \vspace{-4mm}} are finite power series in $g^2$ of order $n$,
which have been fined-tuned to enforce supersymmetry. In
  the   simplest case of the  $SU(2)$ gauge group, the parameters $h_I$  are redundant and can be set to zero. This action permits  us to  perturbatively  compute the renormalized  1PI
generating functional $\Gamma _n$ of the $\varphi_{\dyx}$  at order $n$, such that 
    the Slavnov--Taylor identity   of 
supersymmetry is verified at this order. To obtain the 
action (\ref{baction}) at 
the following order $n+1$, we then use the minimal
subtraction scheme with dimensional reduction
, which defines the
infinite factors $Z_\varphi $ at order $n+1$. They yield as an intermediary result the
``minimally" renormalized 1PI
generating functional $\Gamma_{n+1}^{{\rm \scriptscriptstyle min}}
= \sum_{p=0}^n \Gamma_{{\scriptscriptstyle( p)}} + \Gamma^{{\rm
\scriptscriptstyle min}} _{{\scriptscriptstyle(n+1)}}$. The supersymmetry Slavnov--Taylor identity is possibly  broken at $n+1$ order, as follows\footnote{$\scal{\F,\, \G}$ is the antibracket \vskip - 3mm \petit \be \int d^4 x
\trace \biggl( \frac{\delta^R \F}{\delta A^\mu}
\frac{\delta^L \G}{\delta A^{{\scriptscriptstyle (Q)}}_\mu} + \frac{\delta^R \F}{\delta \lambda}
\frac{\delta^L \G}{\delta \overline{\lambda}^{{\scriptscriptstyle (Q)}}}+ \frac{\delta^R
\F}{\delta \phi^i} \frac{\delta^L \G}{\delta \phi^{{\scriptscriptstyle (Q)}}_i} +
\frac{\delta^R \F}{\delta c}\frac{\delta^L \G}{\delta c^{{\scriptscriptstyle (Q)}}} -
\frac{\delta^R \F}{\delta A^{{\scriptscriptstyle (Q)}}_\mu} \frac{\delta^L \G}{\delta A^\mu} +
\frac{\delta^R \F}{\delta \lambda^{{\scriptscriptstyle (Q)}}} \frac{\delta^L \G}{\delta \overline{\lambda}}- \frac{\delta^R \F}{\delta \phi^{{\scriptscriptstyle (Q)}}_i} \frac{\delta^L \G}{\delta \phi^i} -
\frac{\delta^R \F}{\delta c^{{\scriptscriptstyle (Q)}}}\frac{\delta^L \G}{\delta c}\biggr)
\nonumber \ee \norme}
\be\label{truc}
\Q(\Gamma^{{\rm \scriptscriptstyle min}} _{n+1})= \frac{1}{2}
\sum_{p=1}^n{} \scal{ \Gamma _{{\scriptscriptstyle( p)}} , \Gamma _{{\scriptscriptstyle(n+1- p)}} }
+\Q_{| \Sigma} \Gamma ^{{\rm \scriptscriptstyle min}}_{{\scriptscriptstyle( n+1)}} + \mathcal{O}(g^{2n+4})
\ee
Any given  term in the right-hand side of Eq.~(\ref{truc}) may be non-local, but the sum of these terms is a local functional of fields and sources, as   is warranted by
the quantum action principle. There is no supersymmetry anomaly
\cite{white,shadow} and the consistency relation $\Q _{|\Gamma} \Q
(\Gamma)=0$ implies the existence of the local functional $\Sigma^{{\rm \scriptscriptstyle corr}}_{{\scriptscriptstyle(n+1)}}$ such that $\Q(\Gamma^{{\rm \scriptscriptstyle min}} _{n+1} + \Sigma^{{\rm \scriptscriptstyle corr}}_{{\scriptscriptstyle( n+1)}}) = \mathcal{O}(g^{2n+4})$. Thus  the component of order $n+1$ of the parameters $g_I$, $h_I$, $\x_I$,
${\rm y}_I$ and the matrix $N$ can be  modified in such a way that the resulting 1PI generating functional $\Gamma_{n+1}$ satisfies the supersymmetry Slavnov--Taylor identity at order $n+1$.

The   fine-tuning at order $n+1$ will be    achieved
if a large enough number of relations between 1PI Green functions are
satisfied. They are obtained by suitable differentiations of the supersymmetry Slavnov--Taylor identity. The number of ambiguities removed by the Slavnov--Taylor identity is finite and corresponds to that of  parameters of the action. Thus the relations between the 1PI Green functions only have to be implemented on their renormalization conditions. These relations must be expanded on Lorentz and gauge group invariant tensors.

The antighost Ward identities fix the ambiguities on the
Green functions that contain the antishadow $\bar c$ and the $b$
field. The identities
\begin{gather}
\bigl< \bar c^a(p) \lambda^b_\alpha \bigr> = - i p^\mu \bigl< A^{{{\scriptscriptstyle (Q)}}\,a}_\mu
(p) \lambda^b_\alpha \bigr> \CR
\bigl< \bar c^a(p) \bar c^b \bigr> = \alpha (\overline{\epsilon} \ba p
\epsilon) \delta^{ab} - i p^\mu \bigl<A^{{{\scriptscriptstyle (Q)}}\, a}_\mu (p) \bar c^b
\bigr> + i p^\mu \bigl<A^{{{\scriptscriptstyle (Q)}}\, b}_\mu (-p) \bar c^a \bigr>
\end{gather}
permit to compute the value of ${\rm y}_1$ in function of $\x_1$, and ${\rm
 y}_2$ at the $n+1$ order. 


We  first  use the components of the Slavnov--Taylor 
identity that expresses the  closure  of  the
supersymmetry algebra   at the quantum level
\begin{multline}
\bigl< A^{{{\scriptscriptstyle (Q)}}\, a}_\mu (p) \overline{\lambda}^{\alpha\, c} \bigr>
\bigl< \lambda^{{\scriptscriptstyle (Q)}}_{\alpha\, c} (p) A^b_\nu \bigr> + \bigl< A^{{{\scriptscriptstyle (Q)}}\,
a}_\mu (p) c^c \bigr> \bigl< c^{{\scriptscriptstyle (Q)}}_c (p) A^b_\nu \bigr> +
(\overline{\epsilon} \ba p \epsilon)\delta^{ab} \eta_{\mu\nu} \\* + \bigl<
A_\nu^b (-p) A^c_\sigma \bigr> \bigl< A^{{{\scriptscriptstyle (Q)}}\, \sigma}_c (-p) A^{{{\scriptscriptstyle (Q)}}\,
a}_\mu \bigr> = 0 \end{multline}\vskip -10mm \be
\bigl< c^{{{\scriptscriptstyle (Q)}}\, a}(p) A^c_\mu \bigr> \bigl< A^{{{\scriptscriptstyle (Q)}}\, \mu}_c (p)
\lambda^b_\alpha \bigr> + \bigl< c^{{{\scriptscriptstyle (Q)}}\, a}(p) \phi^c_i \bigr> \bigl<
\phi^{{{\scriptscriptstyle (Q)}}\, i}_c (p) \lambda^b_\alpha \bigr> + \bigl< \lambda^b_\alpha
(-p) \overline{\lambda}^{\beta\, c} \bigr> \bigl< \lambda^{{\scriptscriptstyle (Q)}}_{\beta\,
c}(-p) c^{{{\scriptscriptstyle (Q)}}\, a}\bigr> = 0 \nonumber \ee \vskip -10mm \be
 \bigl< A^{{{\scriptscriptstyle (Q)}}\, a}_\mu (p) \overline{\lambda}^{\alpha\, c} \bigr>
\bigl< \lambda^{{\scriptscriptstyle (Q)}}_{\alpha\, c} (p) \phi^b_i \bigr>+ \bigl< A^{{{\scriptscriptstyle (Q)}}\,
a}_\mu (p) c^c \bigr> \bigl< c^{{\scriptscriptstyle (Q)}}_c (p) \phi^b_i \bigr> + \bigl<
\phi^b_i (-p) \phi^c_j \bigr> \bigl< \phi^{{{\scriptscriptstyle (Q)}}\, j}_c (-p) A^{{{\scriptscriptstyle (Q)}}\, a}_\mu
\bigr> = 0 \nonumber \ee \vskip -10mm \begin{multline}
\bigl< \phi^{{{\scriptscriptstyle (Q)}}\, c}_k (p)
\overline{\lambda}^{\alpha\, d} \bigr> \bigl< \lambda^{{\scriptscriptstyle (Q)}}_{\alpha\, d}
(p) \phi^a_i (k) \phi^b_j \bigr> + \bigl< \phi^{{{\scriptscriptstyle (Q)}}\, c}_k (p)
\phi^a_i(k) c^d \bigr> \bigl< c^{{\scriptscriptstyle (Q)}}_d (p+k) \phi^b_j \bigr>\CR
+ \bigl< \phi^{{{\scriptscriptstyle (Q)}}\, c}_k (p) \phi^b_j(-p-k) c^d \bigr>
\bigl< c^{{\scriptscriptstyle (Q)}}_d (-k) \phi^a_i \bigr> + \bigl< \phi^a_i (k) \phi^b_j
(-p-k) A^d_\mu \bigr> \bigl< A^{{{\scriptscriptstyle (Q)}}\, \mu}_d (-k) \phi^{{{\scriptscriptstyle (Q)}}\, c}_k \bigr>
= 0 \nonumber \end{multline}
These identities imply that the quantities  $\x_I$ are functions of 
only two independent parameters. In turn, both parameters are
determined from the following Slavnov--Taylor identities, which express  the supersymmetry covariance of
physical Green functions 
\be \bigl< A^a_\mu (p) A^c_\nu \bigr> \bigl< A^{{{\scriptscriptstyle (Q)}}\, \nu}_c (p)
\lambda^b_\alpha \bigr> + \bigl< \lambda^b_\alpha (-p)
\overline{\lambda}^{\beta\, c} \bigr> \bigl< \lambda^{{\scriptscriptstyle (Q)}}_{\beta\, c}(-p)
A^a_\mu \bigr> = 0 \nonumber \ee
\vskip -10mm
\begin{multline}
\bigl< \phi^a_i (p) \phi^b_j (k) A^d_\mu \bigr> \bigl< A^{{{\scriptscriptstyle (Q)}}\,
\mu}_d (p+k) \lambda^c_\alpha \bigr> + \bigl< \lambda^c_\alpha (
-p-k) \overline{\lambda}^{\beta\, d} \bigr> \bigl<
\lambda^{{\scriptscriptstyle (Q)}}_{\beta\, d} (-p-k) \phi^a_i(p) \phi^b_j \bigr> \\*+ \bigl<
\lambda^c_\alpha ( -p-k) \phi^a_i (p) \overline{\lambda}^{\beta\, d}
\bigr> \bigl< \lambda^{{\scriptscriptstyle (Q)}}_{\beta\, d} (-k) \phi^b_j \bigr> + \bigl<
\lambda^c_\alpha ( -p-k) \phi^b_j (k) \overline{\lambda}^{\beta\, d}
\bigr> \bigl< \lambda^{{\scriptscriptstyle (Q)}}_{\beta\, d} (-p) \phi^a_i \bigr> = 0
\end{multline}
\vskip -10mm 
\begin{multline}
\bigl< \phi^{a_1}_{i_1} (p_1) \phi^{a_2}_{i_2} (p_2) \phi^{a_3}_{i_3}
(p_3) \phi^c_j \bigr> \bigr< \phi^{{{\scriptscriptstyle (Q)}}\, j}_c (p_1+p_2+p_3)
\lambda^b_\alpha \bigr> \\*+ \sum_{r\in \mathds{Z}_3} \bigl<
\lambda^b_\alpha(-p_1-p_2-p_3) \phi^{a_{1+r}}_{i_{ 1+r}} (p_{1+r})
	\phi^{a_{2+r}}_{i_{ 2+r}} (p_{2+r})
		\overline{\lambda}^{\beta\, c}\bigr> \bigl<
		\lambda^{{\scriptscriptstyle (Q)}}_{\beta\, c} (-p_{3+r})
		\phi^{a_{3+r}}_{i_{3+r}} \bigr> \\*
+ \sum_{r\in\mathds{Z}_3} \bigl< \lambda^b_\alpha(-p_1-p_2-p_3)
\phi^{a_{3+r}}_{i_{3+r}} (p_{3+r}) \overline{\lambda}^{\beta\, c}\bigr> \bigl<
\lambda^{{\scriptscriptstyle (Q)}}_{\beta\, c} (-p_{1+r} - p_{2+r}) \phi^{a_{1+r}}_{i_{ 1+r}} (p_{1+r})
\phi^{a_{2+r}}_{i_{ 2+r}} \bigr> \\*+ \bigl<
\lambda^b_\alpha(-p_1-p_2-p_3)\overline{\lambda}^{\beta\, c} \bigr>
\bigl< \lambda^{{\scriptscriptstyle (Q)}}_{\beta\, c} ( -p_1-p_2-p_3) \phi^{a_1}_{i_1} (p_1)
\phi^{a_2}_{i_2} (p_2) \phi^{a_3}_{i_3} \bigr> = 0 \nonumber
\end{multline} 
It remains to determine the matrix $N$, which is related to the terms  quadratic in the sources. This can be done using the identity
\begin{multline}
\bigl< \lambda^b_\beta (-p) \overline{\lambda}^{\gamma\, c}\bigr>
\bigl< \lambda^{{\scriptscriptstyle (Q)}}_{\gamma\, c} (-p) \overline{\lambda}^{{{\scriptscriptstyle (Q)}}\, \alpha\, a}
\bigr> + \bigl<\overline{\lambda}^{{{\scriptscriptstyle (Q)}}\, \alpha\, a} (p) A^c_\mu \bigr>
\bigl< A^{{{\scriptscriptstyle (Q)}}\, \mu}_c(p) \lambda^b_\beta \bigr>\\* 
+ \bigl<\overline{\lambda}^{{{\scriptscriptstyle (Q)}}\, \alpha\, a} (p) \phi^c_i \bigr>
\bigl< \phi^{{{\scriptscriptstyle (Q)}}\, i}_c(p) \lambda^b_\beta \bigr> +
(\overline{\epsilon} \ba p \epsilon) \delta^{ab} \delta^\alpha_\beta =
0 
\end{multline}

\subsection{Renormalization of  local  observables}
We  must also renormalize the part of the action that is linear in
the sources $u$ of the local observables.  Consider a  set of local operators that   mix together  by renormalisation.  Suppose that each one of these operators can be considered as  the element of an  irreducible  supersymmetry multiplet.  Then,  all the other components of the supersymmetry multiplets will mix by renormalisation with the same matrix  of  anomalous dimensions.    As
for the ordinary Green functions,  non-supersymmetric
counterterms must be perturbatively  computed for enforcing the Ward identities.  
The method of the preceding section can be generalized. We decompose each
source into irreducible representations of $Spin(3,1)\times SU(4)$ and
write  the most general gauge-invariant functional linear in
the sources. 
\begin{multline}
\Upsilon^\flat[u,v] = \int d^4 x \Bigl( Z^\k\, {u^i}_i \frac{1}{2} \trace
\phi^{\flat \, j} \phi^\flat_j + Z^\bps \, u_{ij} \frac{1}{2} \trace
\scal{ \phi^{\flat\, i} \phi^{\flat\, j} - \frac{1}{6} \delta^{ij}
 \phi^{\flat \, k} \phi^\flat_k} + Z^\k_1 \, \overline{u}_i \tau^i \trace
\phi^\flat \lambda^\flat \\*+ Z^\bps_1 \, u^\alpha_i
\trace \scal{ \phi^{\flat\, i} \lambda^\flat - \frac{1}{6} \tau^i
 \phi^\flat \lambda^\flat}
+ Z^\k_2 \, u_{[ijk]} \trace \phi^{\flat \, i}
[ \phi^{\flat\, j} , \phi^{\flat\, k}] + Z^{\k\bps}_2 \, u_{[ijk]} \trace\scal{\frac{1}{3}
 \phi^{\flat\, i} \phi^{\flat\, j} \phi^{\flat k} + \frac{1}{8} \overline{\lambda}^\flat \tau^{ijk}
 \lambda^\flat }\\*+
Z^{\bps\k}_2\, \ub_{ijk} \trace\phi^{\flat \, i}
[ \phi^{\flat\, j} ,\phi^{\flat\, k} ] + Z^\bps_2 \, \ub_{ijk} \trace\scal{\frac{1}{3}
 \phi^{\flat\, i} \phi^{\flat\, j} \phi^{\flat\, k} + \frac{1}{8} \overline{\lambda}^\flat \tau^{ijk}
 \lambda^\flat } + \cdots \Bigr)
\end{multline}
There is  an ambiguity   corresponding to each one of the  renormalization factors $Z^\bullet_I$, to be fixed by the supersymmetry Slavnov--Taylor identity. At a given order, we first perturbatively compute the infinite part of the
renormalization factors. Then
the finite part of the renormalization factors must   be adjusted, as  for ordinary Green functions.

Consider as the simplest cases the Konishi operator $\mathcal{O}^{\rm
 K}\equiv \frac{1}{2} \trace \phi^i \phi_i$
and the $\frac{1}{2}$ BPS operator $\mathcal{O}^{ij}_{\rm C}\equiv
\frac{1}{2}\trace (\phi^i \phi^j - 1/6 \, \delta^{ij} \phi^k \phi_k)$.
The renormalization factors of the first  two  components of the associated 
supermultiplets are related because of the Ward identity
\begin{multline}
[\overline{\epsilon} \ba p \tau^j ]^\alpha \bigl< u_{ij} (p)
\phi^a_k(k) \phi^b_l \bigr> + \bigl< u^\alpha_i (p) \phi^a_k (k)
\overline{\lambda}^{\beta\, c} \bigr> \bigl< \lambda^{{\scriptscriptstyle (Q)}}_{\beta\, c}
(p+k) \phi^b_l \bigr> \\*
+ \bigl< u^\alpha_i (p) \phi^b_l (-p-k)
\overline{\lambda}^{\beta\, c} \bigr> \bigl< \lambda^{{\scriptscriptstyle (Q)}}_{\beta\, c}
(-k) \phi^a_k \bigr> = 0 
\end{multline}
A less simple  example, for which there could be non-supersymmetric couterterms with   a   mixing-matrix, is for the cubic operator $\trace \phi^i [ \phi^j,
\phi^k]$ of the Konishi multiplet. The global symmetries and power
counting allow this operator to mix with $ \trace\scal{\frac{1}{3}
 \phi^{[i} \phi^j \phi^{k]} + \frac{1}{8} \overline{\lambda} \tau^{ijk}
 \lambda }$ belonging to  the $\frac{1}{2}$ BPS multiplet associated to
$\mathcal{O}^{ij}_{\rm C}$. The identity
\begin{multline}
3 [ \overline{\epsilon} \tau^{klm} \tau_j ]^\alpha \bigl< u_{klm} (-p_1-p_2-p_3)
\phi^{a_1}_{i_1}(p_1) \phi^{a_2}_{i_2}(p_2) \phi^{a_3}_{i_3} \bigr> \\*+ 3 [ \overline{\epsilon}\tau_j \tau^{klm} ]^\alpha \bigl< \ub_{klm} (-p_1-p_2-p_3)
\phi^{a_1}_{i_1}(p_1) \phi^{a_2}_{i_2}(p_2) \phi^{a_3}_{i_3} \bigr> 
\\* 
+ \sum_{r\in \mathds{Z}_3} \bigl< u^\alpha_j (-p_1-p_2-p_3
)\phi^{a_{3+r}}_{i_{3+r}}(p_{3+r}) \overline{\lambda}^{\beta\, b}
\bigr> \bigl< \lambda^{{\scriptscriptstyle (Q)}}_{\beta\, b}(-p_{1+r} - p_{2+r})
\phi^{a_{1+r}}_{i_{1+r}}(p_{1+r}) \phi^{a_{2+r}}_{i_{2+r}} \bigr> = 0 
 \nonumber 
\end{multline} 
and the component in   ${[\tau^{mnp }]_{\alpha\beta}} $ of the following one
\begin{multline}
3 [ \overline{\epsilon} \tau^{jkl} \tau_i ]^\gamma \bigl< u_{jkl} (p) \lambda^a_\alpha (k)
\lambda^b_\beta \bigr> + 3 [ \overline{\epsilon}\tau_i \tau^{jkl} ]^\gamma \bigl< \ub_{jkl} (p) \lambda^a_\alpha (k)
\lambda^b_\beta \bigr> \\* 
+ \bigl< u^\gamma_i (p) \lambda^a_\alpha (k) \phi^c_j \bigr> \bigl<
\phi^{{{\scriptscriptstyle (Q)}}\, j}_c (p+k) \lambda^b_\beta \bigr> - \bigl< u^\gamma_i (p)
\lambda^b_\beta (-p-k) \phi^c_j \bigr> \bigl<
\phi^{{{\scriptscriptstyle (Q)}}\, j}_c (-k) \lambda^a_\alpha \bigr>= 0
\end{multline}
permit us  to determine perturbatively the renormalization factors of these operators in function
of those of $\mathcal{O}^{\rm K}$ and $\mathcal{O}^{ij}_{\rm C}$.

In fact, the renormalization factors of   all
the other components of the supermultiplet containing
$\mathcal{O}^{\rm K}$ and $\mathcal{O}^{ij}_{\rm C}$ can be perturbatively computed as a  function of those  of the operators $\mathcal{O}^{\rm K}$ and $\mathcal{O}^{ij}_{\rm C}$.

\subsection{Contact terms}
After computing  the   renormalization  of  one insertion of  ``simple"
physical operators in all   Green functions  of fields  $\varphi_{\dyx}$, we may want to
compute their multicorrelators. The renormalization of these correlation
functions   possibly involves the addition of contact
terms.  Such counterterms cannot be generated     in the minimal
scheme prescription. However,  dimensional reduction breaks
supersymmetry, and we  expect     that   finite contact-counterterms    must be added to  the
action,    for restoring supersymmetry. To compute these possible
counterterms, we write the more general $Spin(3,1)\times
SU(4)$-invariant  action that depends only on the sources $u$, in a polynomial way
\begin{multline}
\Xi[u] = \frac{1}{2}\int d^4 x \Bigl( {\rm z}_1 {u^i}_i {u^j}_j + {\rm
z}_2 u^{ij} u_{ij} - i {\rm z}_3 \overline{u}_i \ba \partial u^i -
i {\rm z}_4 \overline{u}_i \tau^{ij} \ba \partial u_j + {\rm z}_5
u^{[ijk]} \partial^\mu \partial_\mu u_{[ijk]}\\* + {\rm z}_6
u^{\{ijk\}} \partial^\mu \partial_\mu u_{\{ijk\}} + {\rm z}_7
{u_j}^{ji} \partial^\mu \partial_\mu {u^k}_{ki} + {\rm z}_ 8
u^\mu_{ij} \partial^\nu \partial_\nu u^{ij}_\mu + \cdots \ \ \Bigr)
\end{multline}
 The values of  the  renormalization factors ${\rm z_I}$ can then be computed, by imposing the supersymmetry Slavnov--Taylor identity,  order by order 
  in perturbation theory. As before,  it is sufficient to enforce   some identities between
  relevant correlation functions. The simplest identity
\be [\ba p \tau_l \epsilon]_\alpha \bigl< u^{kl} (p) u^{ij}\bigr> +
[\overline{\epsilon} \tau^{\{i}]^\beta \bigl< u^{j\}}_\beta (p)
u^k_\alpha \bigr> = 0\ee
  constrains ${\rm z}_3$ and ${\rm z}_4$ as  functions  of ${\rm
 z}_1$ and ${\rm z}_2$, and so on. In practice,  we have  to define   renormalization conditions for each one of the classes of    superficially divergent   correlation functions that are not related by the supersymmetry Slavnov--Taylor identity. Within a given class, the renormalization conditions  of all   correlation functions are related by      supersymmetry Slavnov--Taylor identities. The non-invariant contact-counterterms can then be perturbatively  computed by perturbatively enforcing these renormalization conditions.

\section*{Acknowledgments}
 We thank  Luis Alvarez-Gaum\'e  and  Raymond Stora    for  good  discussions on the subject. 
\subsection*{Acknowledgments}

This work was partially supported under the contract ANR(CNRS-USAR) \\ \texttt{no.05-BLAN-0079-01}.

\end{document}